\documentclass[runningheads]{llncs}
\usepackage{graphicx}
\usepackage{algorithm, algorithmic}
\usepackage{amsmath}
\usepackage{amsfonts}
\usepackage{mathtools}
\usepackage{appendix}
\usepackage{color}
\usepackage{ulem}

\numberwithin{theorem}{section}
\numberwithin{corollary}{section}
\numberwithin{lemma}{section}
\numberwithin{proposition}{section}
\numberwithin{definition}{section}
\numberwithin{remark}{section}

\begin{document}
\title{Maximizing Modular plus Non-monotone Submodular Functions}
\author{Xin Sun\inst{1} \and
Chenchen Wu \inst{2} \and
Dachuan Xu\inst{1} \and
Yang Zhou \inst{3}\thanks{Corresponding author.}
}
\authorrunning{X. Sun, C. Wu, D. Xu, Y. Zhou}
\titlerunning{Maximization of Submodular and Linear Sum}% Part of RIGHT running header

\institute{
Beijing Institute for Scientific and Engineering Computing, Beijing University of Technology, Beijing 100124, P.R. China  \\
\email{athossun@emails.bjut.edu.cn, xudc@bjut.edu.cn}
\and
College of Sciecne, Tianjin University of Technology, Tianjin 300072, P.R. China \\
\email {wu\_chenchen\_tjut@163.com}
\and
School of Mathematics and Statistics, Shandong Normal University, Jinan 250014, P.R. China\\
\email{zhyg1212@163.com}
}
\maketitle              % typeset the header of the contribution
\begin{abstract}
\label{abstract}
The study of submodular maximization has drawn much attention from both theoretical and practical perspectives, since the objective owns a very general property, i.e., submodularity.
It provides a unified framework for various well-known classical problems in combinatorial optimization, such as max-cut, generalized assignment and facility location. The research problem in this work is the relaxation of maximizing non-negative submodular plus modular with the entire real number domain as its value range over a family of down-closed sets. We seek a feasible point $\mathbf{x}^*$ in the polytope of the given constraint such that  $\mathbf{x}^*\in\arg\max_{\mathbf{x}\in\mathcal{P}\subseteq[0,1]^n}F(\mathbf{x})+L(\mathbf{x})$, where $F$, $L$ denote the extensions of the underlying submodular function $f$ and modular function $\ell$. We provide an approximation algorithm named \textsc{Measured Continuous Greedy with Adaptive Weights}, which yields a guarantee $F(\mathbf{x})+L(\mathbf{x})\geq \left(1/e-\mathcal{O}(\epsilon)\right)\cdot f(OPT)+\left(\frac{\beta-e}{e(\beta-1)}-\mathcal{O}(\epsilon)\right)\cdot\ell(OPT)$ under the assumption that the ratio of non-negative part within $\ell(OPT)$ to the absolute value of its negative part is demonstrated by a parameter $\beta\in[0, \infty]$, where $OPT$ is the optimal integral solution for the discrete problem. It is obvious that the factor of $\ell(OPT)$ is $1$ when $\beta=0$, which means the negative part is completely dominant at this time; otherwise the factor is closed to $1/e$ whe $\beta\rightarrow\infty$. Our work first breaks the restriction on the specific value range of the modular function without assuming non-positivity or non-negativity as previous results and quantifies the relative variation of the approximation guarantee for optimal solutions with arbitrary structure. Moreover, the algorithm we presented still works when the submodular function is monotone and achieves $F(\mathbf{x})+L(\mathbf{x})\geq (1-1/e)f(OPT)+\left(\frac{\beta-e}{e(\beta-1)}-\mathcal{O}(\epsilon)\right)\ell(OPT)$ approximation. Besides, we also give an analysis for the inapproximability of the problem we consider. We show a hardness result that there exists no polynomial algorithm whose output $S$ satisfies $f(S)+\ell(S)\geq0.478\cdot f(OPT)+\ell(OPT)$.

\keywords{Submodular maximization \and Modular function \and Measured continuous greedy \and Down-closed family of sets \and Curvature}
\end{abstract}
\section{Introduction}
\label{sec_introduction}
Recently, plenty of works related to submodular maximization have been published from both theoretical and practical perspectives, since the objective of this problem owns a very general property, i.e., submodularity.
In theory, submodular maximization gives a unified and general framework for many problems in combinatorial optimization such as max-cut \cite{GW95,Hastad01,Karp72,Khot07}, max-dicut \cite{FG95,GW95,HZ01}, max-$k$ coverage \cite{Feige98,KMN99}, max-bisection \cite{ABG16,FJ97}, generalized assignment \cite{CK05,CKR06,FV06,Fleischer06} and facility location \cite{AS99,CFN77a,CFN77b}. 
In practice, the trail of submodular functions can be found from different angles with many applications, for example, social welfare \cite{Von08,GKK20,LV21}, sensor placement \cite{KSG08,MP19}, influence maximization \cite{KKT15,BRS17,GW21}, data privacy \cite{Gupta10,Mitrovic17,RY20,CNZ21} and subset selection \cite{DK08,DK11,Tschiatschek14,NM19,CIS21} in machine learning.

For now, a common paradigm for designing an efficient approximation algorithm of the problem above is continuous greedy based methods plus rounding. 
The reason why it is powerful lies in that it provides a unified framework for solving submodular maximization and its variants instead of being constrained by the the specific structure like combinatorial methods.
By applying the technique above, C$\breve{\rm a}$linescu et al. \cite{CCPV11} presented the first randomized $(1-1/e)$-approximation algorithm for the problem of submodular maximization over a matroid constraint, where the objective is monotone. This result achieved the tight bound given by Nemhauser and Wolsey \cite{NW78} for cardinality constraint, which is the special case of matroid.
The continuous greedy framework contains two stages. It starts with solving the relaxation problem defined by the multilinear extension of the given set objective. Although the multilinear extension is neither concave nor convex \cite{CCPV11}, a feasible fractional solution of the relaxation can still be efficiently computed only if the polytope of the given constraint is solvable. Historically, to obtain an integer feasible solution, pipage \cite{AS04} and swap rounding \cite{CVZ10} are prominent methods since they can get the discrete solutions without loss of approximation.
Moreover, Vondr\'ak \cite{Von10} generalized the guarantee obtained by the continuous greedy approach to $(1-e^{-c}/c)$ with the assumption of bounded curvature $c$. 
This notion is given by Conforti and Cornu\'ejols \cite{CC84}. It measures the distance that a set function deviates from linearity. Specifically, a function is modular when curvature equals to $0$ and in this case the above guarantee tends to $1$.
Furthermore, Sun et al. \cite{SXGL21} showed a fully deterministic algorithm for maximizing a submodular subject to matroid constraints, which also concluded the optimal solution can always be produced when the submodular objective degenerates to an affine one.
Besides, as one of the most classic methods in discrete optimization, local search also makes great contribution in submodular maximization. Filmus and Ward \cite{FW14} offered a $(1-e^{-c})/c$ with restricted curvature. However, the algorithm is specifically designed according to the axioms of a matroid and the Brualdi's lemma \cite{Brualdi69}. 

An interesting observation of both above results is the approximation ratio improves gradually when the curvature tends to $0$, which implies that the set function progressively turns itself into modular. Based on this, Sviridenko et al. \cite{Svi17} improved the guarantee from $(1-e^{-c}/c)$ to $(1-c/e)$. The result is obtained by approximately solving a maximization problem of a non-negative and non-decreasing submodular set function plus a (possibly negative) linear set function. Formally, the following problem needs to be considered
$$\max_{S\in\mathcal{M}}f(S)+\ell(S),$$
where the pair $\mathcal{M}=(\mathcal{S},\mathcal{I})$ denotes the matroid constraint w.r.t. a given ground set $\mathcal{S}$ and the independent system $\mathcal{I}$.
For this problem, Sviridenko et al. \cite{Svi17} described a modified non-oblivious local search and an adapted continuous greedy, respectively. Both algorithms achieved $f(S)+\ell(S)\geq (1-1/e)\cdot f(OPT)+\ell(OPT)$ with $\mathcal{O}(\epsilon)$ error term. Unfortunately, a guessing step is necessary which remarkably damages its query complexity.
Recently, Feldman \cite{Feldman21} considered the multilinear relaxation of the above problem with a slightly more difficult constraint, since there is no lossless rounding techniques. In this work, a clean alternative algorithm is described. It makes the continuous greedy become a guess-free algorithm and whose output $\mathbf{x}\in\mathcal{P}$ obeys a tight guarantee since the ratios of the submodular and modular terms of the objective sums are $(1-1/e)$ and $1$, respectively. Also, an inapproximability is included in it and it states that there exists no polynomial time algorithm with $\mathbb{E}[f(S)+\ell(S)]\geq\max_{\lambda\in[0,1],T\subseteq\mathcal{I}} \{(1-e^{-\lambda}+\epsilon)\cdot f(T)+\lambda\cdot\ell(T)\}$, where $\lambda$ is a calibrating parameter. 

In our paper, a more challenging setting is considered, where the submodular component of the objective is not necessary monotone. In this case, the continuous greedy normally fails, since the marginal gains of the selected direction may not be non-negative any more. For non-monotone submodular maximization, a powerful technique named measured continuous greedy is introduced by Feldman et al. \cite{FNS11} as a unified algorithm. It compensates for the difference between the residual value of the current fractional solution and its gradient. This is achieved by multiplying a factor, which reduces the update rate and distorts the direction with a multiplicative factor, so as to mimic the gradient value. Fortunately, the first-order property of the multilinear extension can ensure this intention. Based on the method above, Lu et al. \cite{LYG21} considered the sums of non-monotone submodular and non-positive modular functions on more specific constraints such as matroid and cardinality with outputting a solution $S$ satisfying $\mathbb{E}\left[f(S)+\ell(S)\right]\geq 1/e\cdot f(OPT)+\ell(OPT)-\mathcal{O}(\epsilon)$. Comparing with their work, we study the maximization of modular plus non-monotone submodular functions. Note that there is no limitation on the modular function $\ell$ like non-positive or non-negative. Also, our constraint is a down-closed family of sets, which is more general and includes some common constraints e.g. matroid, cardinality and knapsack. Although we are all inspired by measured continuous greedy from the algorithm perspective, there are differences in the linear programming sub-problem constructed and the update formula of iteration points. But it is very interesting that both algorithms can get the same approximate guarantee. Furthermore, our result quantifies the relative variation of the approximation guarantee for optimal solutions with arbitrary structure, which means our work generalize \cite{LYG21}. Concretely, we use a parameter to demonstrate the ratio of non-negative part within the optimal value of $\ell$ to the absolute value of its negative part. Then, the variation process of the approximation guarantee for the modular component can be precisely reflected by different values of the parameter.

\subsection{Our Contribution}
\label{contribution}
We first present a near $F(\mathbf{x})+L(\mathbf{x})\geq 1/e\cdot f(OPT)+ \frac{\beta-e}{e(\beta-1)}\cdot\ell(OPT)$ guarantee for maximizing the extensions' sum of $F(\cdot)$ and $L(\cdot)$ over a general solvable down-closed polytope, where $OPT$ denotes the optimal integral solution for the problem. The underlying set functions of the multilinear extension $F(\cdot)$ and $L(\cdot)$ are $f$ and $\ell$, respectively, where $f$ is submodular but non-monotone and the set function $\ell$ is modular. Here the parameter $\beta\in[0, \infty]$ is the ratio of non-negative part within $\ell(OPT)$ to the absolute value of its negative part. Formally, $\beta=\sum_{o\in OPT^+}\ell(o) / |\sum_{o\in OPT^-}\ell(o)|$ and $OPT^+=\{o~|~\ell(o)\geq 0, o\in OPT\}$, $OPT^-=\{o~|~\ell(o)< 0, o\in OPT\}$, respectively.
The variation of $\beta$ reflects the change process of the approximation guarantee of the modular term in the sums. When $\beta=0$, which means the negative part of $OPT$ is completely dominant in $\ell(OPT)$, the factor of $\ell(OPT)$ is $1$. When the positive part mainly controls $\ell(OPT)$, i.e., $\beta\rightarrow\infty$, then we obtain near $1/e$ for the modular component. Our algorithm is \textsc{Measured Continuous Greedy with Adaptive Weights}. 
In each round, it solves a linear programming with a convex constraint in polynomial time. The weight in the objective is adaptive, and it increases when the timestamp moves forward. This indicates we pay more attention to the modular function initially and the other part is more and more valued in the solution with the growth of time. Finally, they are equally valued in the last round.
The outputs of these adaptive processes lead to the update direction of the current fractional solution in next step. The informal theorem we obtained is given below.

\begin{theorem}{(Informal)}
	\label{thm_informal}
	The \textsc{Measured Continuous Greedy with Adaptive Weights} outputs $\mathbf{x}\in\mathcal{P}$ and with high probability, it satisfies 
	$$F(\mathbf{x})+L(\mathbf{x})\geq (1/e-\epsilon)\cdot f(OPT)+\left(\frac{\beta-e}{e(\beta-1)}-\mathcal{O}(\epsilon)\right)\cdot\ell(OPT)-\mathcal{O}(\epsilon).$$
\end{theorem}
 
Moreover, the algorithm that we use also is suitable for monotone case. We show it can achieve the same ratio as Feldman did in \cite{Feldman21}. 

\begin{theorem}{(Informal)}
	\label{thm_monotone_informal}
	When $f$ is monotone, algorithm \textsc{Measured Continuous Greedy with Adaptive Weights} can produce a vector $\mathbf{x}\in\mathcal{P}$ and with high probability it satisfies 
	$$F(\mathbf{x})+L(\mathbf{x})\geq \max_{\lambda\in[0,1]}(\lambda/e-\epsilon)\cdot f(OPT)+\lambda\left(\frac{\beta-e}{e(\beta-1)}-\mathcal{O}(\epsilon)\right)\cdot\ell(OPT).$$
\end{theorem}

Besides, inspired by the conclusion given in \cite{GV11}, we present the hardness result for maximizing non-monotone and non-negative submodular plus modular functions subject to a down-closed family of sets.

\begin{theorem}
	\label{hardness_SumSubMod_informal}
	For the problem of maximizing non-monotone and non-negative submodular plus modular functions subject to a down-closed family of sets, no algorithm can output a feasible solution $S$ with  
	$$\mathbb{E}[f(S)+\ell(S)]\geq\max_{\lambda\in[0,1],T\subseteq\mathcal{I},|T|\leq k}\{0.478\cdot g(T)+\lambda\cdot \ell(T)\}$$ 
	in polynomial time, where $k$ denote the size of the maximal feasible set in $\mathcal{I}$.
\end{theorem}

\subsection{Organization}
\label{organization}
%
% the environments 'definition', 'lemma', 'proposition', 'corollary',
% 'remark', and 'example' are defined in the LLNCS documentclass as well.
%
The remainder of the paper is organized as below. In Section \ref{sec_preliminary}, we discuss the basic notations including the multilinear extension and its significant properties, which makes contribution for designing our algorithms. Moreover, some technical lemmata are introduced in order to help us finish the analysis. In Section \ref{sec_analysis}, the algorithm \textsc{Measured Continuous Greedy with Adaptive Weights} is showed and its analysis is divided into three parts. In Section \ref{sec_addtionalresults}, we briefly present some additional results obtained by our algorithms. Besides, we also present an inapproximability result in Section \ref{sec_hardness}. Finally, Section \ref{sec_conclusion} gives a conclusion.

\section{Preliminaries}
\label{sec_preliminary}
Given a ground set $\mathcal{S}$ with $n$ elements and a set function $f: 2^{\mathcal{S}}\rightarrow\mathbb{R}_{\geq 0}$, we use $f_S(\{s\})\coloneqq f(S\cup\{s\})-f(S)$ to denote the marginal gains of adding an element $s$ to a set $S\in\mathcal{S}$ w.r.t. $f$. From the perspective of marginal gains, we say $f$ is submodular if $f_S(\{s\})\geq f_T(\{s\})$ holds for any $S\subseteq T\subseteq \mathcal{S}$ and an element $s\in \mathcal{S}\backslash T$. Also, the set function $f$ is non-negative if $f(S)\geq 0$ for any $S\subseteq \mathcal{S}$. Besides, a set function $\ell: 2^{\mathcal{S}}\rightarrow\mathbb{R}$ is modular if it is not only submodular but also supermodular (and a function is supermodular if its negativity is submodular). An important characteristic for such $\ell$ is that it is additive, i.e., $\ell(S)=\sum_{s\in S}\ell(\{s\})$. For easy of presentation, we omit the brace when adding an element and define $S\cup\{s\}$ and $S\backslash\{s\}$ by the shorthand $S\cup s$ and $S-s$, respectively. Since the representation of a submodular function might be $\mathcal{O}(e^n)$, we make an assumption that for any query $f(S)$ of a set $S\subseteq\mathcal{S}$, there is a value oracle, which gives response to it immediately.

In this paper, we require that the solutions are feasible to the down-closed family of subsets $\mathcal{I}\subseteq 2^{\mathcal{S}}$, which is a quite general constraint including several regular and natural constraints such as matroid and knapsack in submodular maximization. $\mathcal{I}$ is down-closed (or down-monotone) if for any feasible set in $\mathcal{I}$, its all subsets also belong to $\mathcal{I}$. Since we indeed aim at the relaxation problem, we refer to $\mathcal{P}\in[0, 1]^n$ as the feasible polytope of $\mathcal{I}$. It can be constructed by the convex hull of the points w.r.t. all feasible sets in $\mathcal{I}$. We say $\mathcal{P}$ is down-monotone if $0\leq\mathbf{y}\leq\mathbf{x}$ and $\mathbf{x}\in\mathcal{P}$ imply $\mathbf{y}\in\mathcal{P}$. Moreover, if linear functions can be maximized over the polytope $\mathcal{P}$ in polynomial time, we call $\mathcal{P}$ solvable.
In this paper, the boldface represents vectors and we use $\mathbf{x}_0$, $\mathbf{x}_1$ with different subscripts to distinguish vectors. Besides, for a vector $\mathbf{x}\in[0,1]^n$, its components are defined by  $\{\mathbf{x}(s)\}_{s\in\mathcal{S}}$.

For a set function $f$, its multilinear extension is defined as $F: [0,1]^n\rightarrow\mathbb{R}_{\geq 0}$, which constructs a mapping between a point $\mathbf{x} \in [0,1]^{n}$ and the expected function value of a random set $R_{\mathbf{x}}\subseteq\mathcal{S}$. For the construction of the set $R_{\mathbf{x}}$, each element $s\in\mathcal{S}$ is included in it with probability $\mathbf{x}(s)$ independently. The formal mathematical expression is $F(\mathbf{x}) \coloneqq \mathbb{E}[f(R_{\mathbf{x}})] = \sum_{S \subseteq \mathcal{S}}f(S) \prod_{s\in S}\mathbf{x}(s) \prod_{s\notin S}(1-\mathbf{x}(s))$.
We denote the coordinate-wise maximum and minimum of $\mathbf{x}$ and $\mathbf{y}$ by $\mathbf{x}\vee\mathbf{y}(s)=\max\{\mathbf{x}(s), \mathbf{y}(s)\}$ and $\mathbf{x}\wedge\mathbf{y}(s)=\min\{\mathbf{x}(s), \mathbf{y}(s)\}$ for any two vectors $\mathbf{x}, \mathbf{y} \in [0,1]^{n}$, respectively.
Now we can finally give a description of the problem. We intend to search for a point $\mathbf{x}\in\mathcal{P}$ maximizing the relaxation $F(\mathbf{x})+L(\mathbf{x})$, where $L(\cdot)$ is the modular extension of $\ell$. It is actually the dot product of the input and $\vec{\ell}$, where $\vec{\ell}$ denotes the vector whose components are the coordinate-wise value of the modular function. Formally, $L(\mathbf{x})\coloneqq\langle\vec{\ell},\mathbf{x}\rangle=\sum_{s\in\mathcal{S}}\ell(s)\cdot\mathbf{x}(s)$.
Therefore, it is naturally that $L(\mathbf{1}_S)=\ell(S)$ for all subsets $S\in\mathcal{S}$. 
Let $OPT$ be the set corresponding to an optimal integral solution for the problem, i.e., $OPT=\arg\max_{S\subseteq\mathcal{S}}\{f(S)+\ell(S)\}$, and let $\tau=\max_{s\in\mathcal{S}}f(s)$ be the largest value of a single element in the ground set.

Furthermore, we give a few useful and technical lemmata before analysis. The first one is a powerful lemma that was first proved in \cite{AS00}. For several mutually independent random variables, it shows the upper bound probability of their sums beyond a certain value.
\begin{lemma}{(\cite{AS00})}
	\label{tech_lem_propstat}
	For mutually independent random variables $X_i$, $1\leq i\leq k$, let all $|X_i|\leq 1$ and $\mathbb{E}[X_i]=0$. Then $\Pr[|\sum_{i=1}^{k}X_i|>a]\leq 2e^{-a^2/2k}$.
\end{lemma}

Next, we give some known structural lemmata about multilinear extension. The first is the first-order property, which shows several equivalent forms about the derivative in a certain dimension.
\begin{lemma}
	\label{multilinear_first-order}
	For a submodular set function $f: 2^{\mathcal{S}}\rightarrow\mathbb{R}_{\geq 0}$, $F(\mathbf{x})$ is the multilinear extension of it. For every element $s\in\mathcal{S}$, we have the following equalities as the first-order properties
	\begin{eqnarray*}
		\partial_sF(\mathbf{x})
		&=&\frac{F(\mathbf{x}\vee\mathbf{1}_s)-F(\mathbf{x})}{1-\mathbf{x}_s} \\
		&=&\frac{F(\mathbf{x})-F(\mathbf{x}\wedge\mathbf{1}_{\mathcal{S}-s})}{\mathbf{x}_s} \\
		&=&F(\mathbf{x}\vee\mathbf{1}_s) - F(\mathbf{x}\wedge\mathbf{1}_{\mathcal{S}-s}) \\
		&=&\mathbb{E}[f_{R_{\mathbf{x}}-s}(s)].
	\end{eqnarray*}
\end{lemma}
\begin{proof}
	The first two equalities are trivial due the definition of the derivatives.
For the third one, from the definition of multilinear extension we can get
	$$\partial_sF(\mathbf{x}) = \partial\left[\sum_{S \subseteq \mathcal{S}}f(S)\prod_{s^{\prime}\in S}\mathbf{x}(s^{\prime})\prod_{s^{\prime}\notin S}(1-\mathbf{x}(s^{\prime}))\right] / \partial\mathbf{x}(s).$$
	
	Naturally, elements $s$ are either in set $S$ or not in $S$. Then, we have
	\begin{eqnarray*}
		\partial_sF(\mathbf{x})
		&=& \sum_{S: s\in S}f(S)\prod_{s^{\prime}\in S-s}\mathbf{x}(s^{\prime})\prod_{s^{\prime}\notin S}(1-\mathbf{x}(s^{\prime})) \\
		&& - \sum_{S: s\notin S}f(S)\prod_{s^{\prime}\in S}\mathbf{x}(s^{\prime})\prod_{s^{\prime}\notin S\cup s}(1-\mathbf{x}(s^{\prime})) \\
		&=& \sum_{S \subseteq \mathcal{S}}f(S)\prod_{s^{\prime}\in S}(\mathbf{x}\vee\mathbf{1}_s)(s^{\prime})\prod_{s^{\prime}\notin S}\left(1-(\mathbf{x}\vee\mathbf{1}_s)(s^{\prime})\right) \\
		&& - \sum_{S \subseteq \mathcal{S}}f(S)\prod_{s^{\prime}\in S}(\mathbf{x}\wedge\mathbf{1}_{\mathcal{S}-s})(s^{\prime})\prod_{s^{\prime}\notin S}\left(1-(\mathbf{x}\wedge\mathbf{1}_{\mathcal{S}-s})(s^{\prime})\right).
	\end{eqnarray*}
	
	Finally, due to the definition of the multilinear extension, we obtain 
	$$\partial_sF(\mathbf{x})=F(\mathbf{x} \vee \mathbf{1}_s) - F(\mathbf{x} \wedge \mathbf{1}_{\mathcal{S}-s})=\mathbb{E}[f_{R_{\mathbf{x}}-s}(s)].$$
	\qed
\end{proof}

The multilinear nature yields the next lemma. It uses the partial derivatives to approximate the local linearity of the function $F$.
\begin{lemma}{(\cite{FNS11})}
	\label{multilinear_locallinear}
	Let $F(\mathbf{x})$ be the multilinear extension of a submodular function $f: 2^{\mathcal{S}}\rightarrow\mathbb{R}_{\geq 0}$. Then, for any $\mathbf{y}, \mathbf{x}\in[0,1]^{n}$ such that $0\leq \mathbf{y}(s) - \mathbf{x}(s) \leq \delta \leq 1$ for any $s\in\mathcal{S}$, we have
	\begin{eqnarray*}
		F(\mathbf{y}) - F(\mathbf{x}) \geq \sum_{s\in\mathcal{S}}\left(\mathbf{y}(s) - \mathbf{x}(s)\right)\cdot \partial_sF(\mathbf{x}) - n^2\delta^2\cdot\max_{s\in\mathcal{S}}f(s).
	\end{eqnarray*}
\end{lemma}

The last lemma is powerful since it can lower bound the value of any vector joining with a target set as long as the components have a common upper bound.
\begin{lemma}
	\label{multilinear_lowerbound}
	For a submodular set function $f: 2^{\mathcal{S}}\rightarrow\mathbb{R}_{\geq 0}$, $F(\mathbf{x})$ is the multilinear extension of it. For any vector $\mathbf{x}\in[0,1]^{n}$, set $\mathbf{x}(s)\leq\theta$ for any element $s\in\mathcal{S}$. Then, for any set $S\subseteq\mathcal{S}$, we have $F(\mathbf{x}\vee\mathbf{1}_S)\geq(1-\theta)\cdot f(S)$.
\end{lemma}

We omit its proof for now since it can be proved with the help of the Lov\'{a}sz extension of the set function.

\section{Analysis}
\label{sec_analysis}
At the beginning of the section, we show the Measured Continuous Greedy with Adaptive Weights (stated as Algorithm \ref{MCGAW}) for the submodular maximization of the multilinear extension $F(\cdot)$ of a non-negative submodular function plus the extension of a modular function $\ell(\cdot)$ without any limitation under the constraint of a solvable and down-monotone polytope $\mathcal{P}$. Algorithm \ref{MCGAW} is inspired by the Measure Continuous Greedy, which is introduced by Feldman et al. \cite{FNS11} as one of the most popular paradigms for non-monotone submodular maximization.
Similar to other continuous greedy based approaches, it begins with $\mathbf{x}_0=\mathbf{1}_{\emptyset}$ and enlarge a solution $\mathbf{x}_t$ slowly over time $t$ until $t=1$.
In each round $t\in[1/\delta]$, Algorithm \ref{MCGAW} solves a linear programming in polynomial time and outs a vector $\mathbf{z}_t$. The weight in the objective is adaptive, and it increases when the timestamp $t$ moves forward. This means we pay more attention to the modular function initially and the other part is more and more valued in the solution with the growth of time. Finally, they are equally valued until $t=1$.
The outputs $\{\mathbf{z}_t\}$ of these adaptive processes lead to the update direction $\mathbf{z}_t\odot(1-\mathbf{x}_t)$ of the current fractional solution $\mathbf{x}_t$ in next step, where $\odot$ defines the hadamard product of two vectors. Note that the multiplicative factor of $(1-\mathbf{x}_t)$ can be seen as a distinguished symbol of measured continuous greedy technique, since it remedies the difference between the gradient of the current solution $\mathbf{x}$ and its residual to the optimal value. As for the gradient, we use its good enough estimation $w_{\mathbf{x}}$ since the computation of multilinear extension requires exponentially queries to the value oracle. 

\begin{algorithm}[htbp]
	\caption{\textsc{Measured Continuous Greedy with Adaptive Weights}}
	\label{MCGAW}
	\begin{algorithmic}[1]
		\REQUIRE submodular function $f$, modular vector $\vec{\ell}$, polytope $\mathcal{P}$, parameter $\epsilon$.
		\ENSURE $\mathbf{x}_1$
		\STATE Set: $\mathbf{x}_0\leftarrow\mathbf{1}_{\emptyset}$, $t\leftarrow0$, $\delta\leftarrow\lceil2n^2/\epsilon\rceil^{-1}$.
		\WHILE {$t < 1$}
		\STATE draw $d=\lceil \frac{n^2\ln(n^2/\delta)}{2e^2} \rceil$ independently sets $\{R^j_{\mathbf{x}_t}\}_{j=1}^d$ such that each element $s\in\mathcal{S}$ belongs to set $R^j_{\mathbf{x}_t}$ with probability $\mathbf{x}_t(s)$.
		\FOR {$s\in\mathcal{S}$}
		\STATE $w_{\mathbf{x}_t}(s)\leftarrow\frac{1}{d}\sum_{j=1}^df_{R^j_{\mathbf{x}_t}-s}(s)$.
		\ENDFOR
		\STATE $\mathbf{z}_{t}\leftarrow\arg\max_{\mathbf{z}_{t}\in \mathcal{P}}\left\langle \mathbf{z}_{t}\odot(1-\mathbf{x}_{t}), (1+\delta)^{(t-1)/\delta}\cdot w_{\mathbf{x}_t}+   \vec{\ell} \right\rangle$.
		\STATE $\mathbf{x}_{t+\delta}\leftarrow \mathbf{x}_{t}+\delta\cdot \mathbf{z}_{t}\odot(1-\mathbf{x}_{t})$.
		\STATE $t\leftarrow t+\delta$.
		\ENDWHILE
	\end{algorithmic}
\end{algorithm}

\subsection{Function Estimation}
From the first-order property (stated as Lemma \ref{multilinear_first-order}) of multilinear extension, we know the derivative is actually the expectation of the marginal gain. In this subsection we show that it can be estimated well by sampling.
We consider the event $\Xi$ that $|w_{\mathbf{x}_t}(s)-\mathbb{E}[f_{R_{\mathbf{x}_t}-s}(s)]|\leq2\epsilon\tau/n$ for every element $s\in\mathcal{S}$ and every timestamp $t$ in Algorithm \ref{MCGAW}, where $\tau=\max_{s\in\mathcal{S}}f(s)$. 
For the convenience of analysis, we denote $\mathcal{G}=\{k\delta | k\in\mathbb{Z}_{+}, 0\leq k <\delta^{-1}\}$ be the set of timestamps considered by Algorithm \ref{MCGAW}, since we have to do the discretization to a continuous method on a discrete computer. 
The lemma in this subsection shows that the event $\Xi$ happens with high probability, which means that if the sampling error can be well bounded, then $w_{\mathbf{x}}(s)$ obtained in Algorithm \ref{MCGAW} is normally a good proxy for $\partial_sF(\mathbf{x})$.

\begin{lemma}
	\label{func_esti}
	For any round $t\in\mathcal{G}$ in Algorithm \ref{MCGAW}, we assume the event $\Xi$ denotes $|w_{\mathbf{x}_t}(s)-\mathbb{E}[f_{R_{\mathbf{x}_t}-s}(s)]|\leq2\epsilon\tau/n$ for each element $s\in\mathcal{S}$, where $\tau=\max_{s\in\mathcal{S}}f(s)$. Then $\Xi$ happens with probability $\mathbb{P}[\Xi]\geq 1-\frac{2}{n}$.
\end{lemma}
\begin{proof}
	Recall that for each pair $(t,s)\in\mathcal{G}\times\mathcal{S}$, we calculate the estimation $w_{\mathbf{x}_t}(s)$, which approximates the expectation value $\mathbb{E}[f_{R_{\mathbf{x}_t}-s}(s)]$ of adding element $s$ to a random set $R_{\mathbf{x}_t}$ with respect to the current fractional solution $\mathbf{x}_t$.
	For this calculation, we obtain a independent set sequence $\{R^j_{\mathbf{x}_t}\}_{j=1}^d$ by $d$ times sampling.
	Each random set $R^j_{\mathbf{x}_t}$ in it is drawn by doing $n$ times Bernoulli trials $\mathcal{B}({\mathbf{x}_t(s)})$ with different “success" probabilities in $\mathbf{x}_t(s)$. 
	Now, we consider the random variable sequence $\{Y^j_{\mathbf{x}_t}\}_{j=1}^{d}$ with definition
	$$Y^j_{\mathbf{x}_t} = \dfrac{f_{R^j_{\mathbf{x}_t}-s}(s)-\mathbb{E}[f_{R_{\mathbf{x}_t}-s}(s)]}{2\tau}.$$
	
	It is obviously $\mathbb{E}[Y^j_{\mathbf{x}_t}]=0$ as a result of the linearity of the expectation and $f_{R^j_{\mathbf{x}_t}-s}(s)\leq\tau$ (also true for $\mathbb{E}[f_{R_{\mathbf{x}_t}-s}(s)]$) due to the submodularity of $f$. Consequently, we can get $Y^j_{\mathbf{x}_t}\in[-1,1]$, whereas $f_{R^j_{\mathbf{x}_t}-s}(s)$ and $\mathbb{E}[f_{R_{\mathbf{x}_t}-s}(s)]$ might be negative due to the non-monotonicity of $f$.
	
	Since $w_{\mathbf{x}_t}(s)$ is obtained by averaging the marginal value with respect to the $d$ samplings, we can get
	$$\left|w_{\mathbf{x}_t}(s)-\mathbb{E}\left[f_{R_{\mathbf{x}_t}-s}(s)\right]\right|=\frac{2\tau}{d}\cdot\left|\sum_{j=1}^{d}Y^j_{\mathbf{x}_t}\right|.$$
	
	Then, due to the Theorem A.1.16 in \cite{AS00} (stated as Lemma \ref{tech_lem_propstat}), we can upper bound the probability, where the event $\Xi$ with respect to a single pair $(t,s)$ does not happen.
	\begin{eqnarray*}
		\mathbb{P}\left[|w_{\mathbf{x}_t}(s)-\mathbb{E}(f_{R_{\mathbf{x}_t}-s}(s))|>\frac{2\epsilon\tau}{n}\right] 
		&=& \mathbb{P}\left[\frac{2\tau}{d}\cdot\left|\sum_{j=1}^{d}Y^j_{\mathbf{x}_t}\right|>\frac{2\epsilon\tau}{n}\right] \\
		&=& \mathbb{P}\left[\left|\sum_{j=1}^{d}Y^j_{\mathbf{x}_t}\right|>\frac{d\epsilon}{n}\right] \\
		&\leq& 2e^{-(\frac{2d\epsilon}{n})^2/2d} = 2e^{-2d\epsilon^2/n^2} \\
		&\leq& 2e^{\ln(\delta/n^2)} = \frac{2\delta}{n^2}.
	\end{eqnarray*}
	
	By the union bound, the probability that there exists a pair $(t,s)\in\mathcal{G}\times\mathcal{S}$ with 
	$$|w_{\mathbf{x}_t}(s)-\mathbb{E}(f_{R_{\mathbf{x}_t}-s}(s))|>\frac{\epsilon\tau}{n}$$
	is at most $|\mathcal{S}|\cdot|\mathcal{G}|\cdot\frac{2\delta}{n^2}=\frac{2}{n}$, since $|\mathcal{G}|=\delta^{-1}$ and $|\mathcal{S}|=n$. Thus, event $\Xi$ happens with high probability $\mathbb{P}[\Xi]=1-\frac{2}{n}$.	\qed
\end{proof}

\subsection{Auxiliary Function}
Recall that the weights in the objective of the linear programming in Algorithm \ref{MCGAW} are adaptive. 
The objective measures the increase on the sums of one iterative round. As it presents, the weight of the modular function remains unchanged at $1$, while the weight of the gradient estimation varies itself with time from $(1+\delta)^{-1/\delta}$ to $1$. Therefore, we  can denote an auxiliary function $\Gamma(t)\coloneqq(1+\delta)^{(t-1)/\delta}\cdot F(\mathbf{x}_t)+L(\mathbf{x}_t)$ in order to lower bound the update revenue.
Our next lemma can be viewed as a central component in the later proofs, which claims the relationship between the update profit per unit time and the residual value with respect to the current solution $\mathbf{x}_t$.

\begin{lemma}
	\label{update_profit}
	With probability $1-\mathcal{O}(\frac{1}{n})$, for all $t\in\mathcal{G}$, 
	\begin{eqnarray*}
		\dfrac{\Gamma(t+\delta)-\Gamma(t)}{\delta}&\geq& (1+\delta)^{(t-1)/\delta}\left[F(\mathbf{x}_t\vee\mathbf{1}_{OPT})+F(\mathbf{x}_{t+\delta})-F(\mathbf{x}_t)\right]\\
		&+& \sum_{o\in OPT}\ell(o)(1-\mathbf{x}_t(o)) -5\epsilon\tau.
	\end{eqnarray*}
\end{lemma}
\begin{proof}
	We condition on the event $\Xi$ for the rest of the proof. 
	Note that $\mathbf{z}_t\odot(1-\mathbf{x}_t)\in\mathcal{P}\subseteq[0,1]^n$, since $\mathbf{z}_t$ and $(1-\mathbf{x}_t)$ are both in $\mathcal{P}$ and $\mathcal{P}$ is down-monotone.
	Then, we can get 
	\begin{eqnarray*}
		F(\mathbf{x}_{t+\delta}) - F(\mathbf{x}_t) &\geq& \sum_{s\in\mathcal{S}}(\mathbf{x}_{t+\delta}-\mathbf{x}_t)\cdot\partial_sF(\mathbf{x}_t) - n^2\delta^2\cdot\max_{s\in\mathcal{S}}f(s) \\
		&\geq& \sum_{s\in\mathcal{S}}\delta \mathbf{z}_t(s)(1-\mathbf{x}_t(s))\cdot\mathbb{E}[f_{R_{\mathbf{x}_t}-s}(s)]-\epsilon\delta\tau \\
		&\geq& \sum_{s\in\mathcal{S}}\delta \mathbf{z}_t(s)(1-\mathbf{x}_t(s))\cdot [w_{\mathbf{x}_t}(s)-2\epsilon\tau/n] - \epsilon\delta\tau \\
		&\geq& \langle \delta \mathbf{z}_t\odot(1-\mathbf{x}_t), w_{\mathbf{x}_t}\rangle - 3\epsilon\delta\tau,
	\end{eqnarray*}
	where the first inequality holds by the local nearly linearity of multilinear extension (stated as Lemma \ref{multilinear_locallinear}), the second inequality follows from the definition of $\delta$ since $\delta\leq\epsilon n^{-2}$, the third one is true since we assume that the event $\Xi$ happened and the final one is given by $\mathbf{z}_t(s)\cdot(1-\mathbf{x}_t(s))\leq1$ for each $s\in\mathcal{S}$.
	
	Due to the definition of $\Gamma$ and the additivity of $\ell$, we further get
	\begin{eqnarray*}
		\dfrac{\Gamma(t+\delta)-\Gamma(t)}{\delta}&=&\dfrac{(1+\delta)^{(t+\delta-1)/\delta}\cdot F(\mathbf{x}_{t+\delta})-(1+\delta)^{(t-1)/\delta}\cdot F(\mathbf{x}_t)}{\delta} \\
		&&+~ \dfrac{L(\mathbf{x}_{t+\delta})-L(\mathbf{x}_t)}{\delta}\\
		&=& \dfrac{(1+\delta)^{(t-1)/\delta}\cdot [F(\mathbf{x}_{t+\delta})-F(\mathbf{x}_t)]}{\delta} \\
		&&+~ (1+\delta)^{(t-1)/\delta}\cdot F(\mathbf{x}_{t+\delta}) + \langle \vec{\ell}, \mathbf{z}_t\odot(1-\mathbf{x}_t) \rangle \\
		&\geq& \dfrac{(1+\delta)^{(t-1)/\delta}\cdot [\langle \delta \mathbf{z}_t\odot(1-\mathbf{x}_t), w_{\mathbf{x}_t}\rangle - 3\epsilon\delta\tau]}{\delta} \\
		&&+~ (1+\delta)^{(t-1)/\delta}\cdot F(\mathbf{x}_{t+\delta}) + \langle \vec{\ell}, \mathbf{z}_t\odot(1-\mathbf{x}_t) \rangle \\
		&=& \langle \mathbf{z}_t\odot(1-\mathbf{x}_t), (1+\delta)^{(t-1)/\delta}\cdot w_{\mathbf{x}_t} + \vec{\ell} \rangle\\
		&&+~ (1+\delta)^{(t-1)/\delta}\cdot F(\mathbf{x}_{t+\delta}) - (1+\delta)^{(t-1)/\delta}3\epsilon\tau \\
		&\geq& \langle \mathbf{z}_t\odot(1-\mathbf{x}_t), (1+\delta)^{(t-1)/\delta}\cdot w_{\mathbf{x}_t} + \vec{\ell} \rangle\\
		&&+~ (1+\delta)^{(t-1)/\delta}\cdot F(\mathbf{x}_{t+\delta}) - 3\epsilon\tau,
	\end{eqnarray*}
	where the first inequality holds by the lower bound of $F(\mathbf{x}_{t+\delta})-F(\mathbf{x}_t)$ and the second one is true by observing that $(1+\delta)^{(t-1)/\delta}\leq 1$ since $(t-1)/\delta\leq 0$.
	
	Now we need to focus on the dot product term and use the residual value to lower bound it.
	Recall that $\mathbf{z}_t$ is the vector in the polytope $\mathcal{P}$ maximizing the dot product $\langle \mathbf{z}_t\odot(1-\mathbf{x}_t), (1+\delta)^{(t-1)/\delta}\cdot w_{\mathbf{x}_t}+L(\mathbf{x}_t) \rangle$. Since $\mathbf{1}_{OPT}\in\mathcal{P}$, we get
	\begin{eqnarray*}
		&& \langle \mathbf{z}_t\odot(1-\mathbf{x}_t), (1+\delta)^{(t-1)/\delta}\cdot w_{\mathbf{x}_t}+ \vec{\ell} \rangle \\
		&\geq& \langle \mathbf{1}_{OPT}\odot(1-\mathbf{x}_t), (1+\delta)^{(t-1)/\delta}\cdot w_{\mathbf{x}_t}+ \vec{\ell} \rangle \\
		&=& (1+\delta)^{(t-1)/\delta}\cdot \sum_{o\in {OPT}}w_{\mathbf{x}_t}(o){(1-\mathbf{x}_t(o))} + \sum_{o\in OPT}\ell(o)(1-\mathbf{x}_t(o)) \\
		&\geq& (1+\delta)^{(t-1)/\delta}\cdot \sum_{o\in {OPT}}\left|\mathbb{E}[f_{R_{\mathbf{x}_t}-o}(o)]-2\epsilon\tau/n\right| (1-\mathbf{x}_t(o)) \\
		&&+~ \sum_{o\in OPT}\ell(o)(1-\mathbf{x}_t(o)) \\
		&\geq& (1+\delta)^{(t-1)/\delta}\cdot \sum_{o\in {OPT}}\mathbb{E}[f_{R_{\mathbf{x}_t}-o}(o)](1-\mathbf{x}_t(o)) - 2\epsilon\tau + \sum_{o\in OPT}\ell(o)(1-\mathbf{x}_t(o)) \\
		&=& (1+\delta)^{(t-1)/\delta}\cdot \sum_{o\in {OPT}}[F(\mathbf{x}_t\vee\mathbf{1}_o)-F(\mathbf{x}_t\wedge\mathbf{1}_{\mathcal{S}-o})](1-\mathbf{x}_t(o)) - 2\epsilon\tau \\
		&&+~ \sum_{o\in OPT}\ell(o)(1-\mathbf{x}_t(o)) \\
		&=& (1+\delta)^{(t-1)/\delta}\cdot \sum_{o\in {OPT}}[F(\mathbf{x}_t\vee\mathbf{1}_o)-F(\mathbf{x}_t)] + \sum_{o\in OPT}\ell(o)(1-\mathbf{x}_t(o)) - 2\epsilon\tau \\
		&\geq& (1+\delta)^{(t-1)/\delta}\cdot \left[F(\mathbf{x}_t\vee\mathbf{1}_{OPT})-F(\mathbf{x}_t)\right] + \sum_{o\in OPT}\ell(o)(1-\mathbf{x}_t(o)) - 2\epsilon\tau, \\
	\end{eqnarray*}
	where the second inequality holds since we assume that the event $\Xi$ happened, the third inequality is true by $|OPT|\leq n$, and the final one follows from the submodularity of $f$.
	
	Combining with the above result, we obtain
	\begin{eqnarray*}
		\dfrac{\Gamma(t+\delta)-\Gamma(t)}{\delta}
		&\geq&(1+\delta)^{(t-1)/\delta}\cdot \left[F(\mathbf{x}_t\vee\mathbf{1}_{OPT})+F(\mathbf{x}_{t+\delta})-F(\mathbf{x}_t)\right] \\
		&&+~ \sum_{o\in OPT}\ell(o)(1-\mathbf{x}_t(o)) - 5\epsilon\tau.
	\end{eqnarray*} \qed
\end{proof}

\subsection{Approximation guarantee}
Now, it is time to present the final analysis to the guarantee of Algorithm \ref{MCGAW}. 
By Lemma \ref{update_profit}, the lower bound on the increase of each update is
\begin{eqnarray*}
	\dfrac{\Gamma(t+\delta)-\Gamma(t)}{\delta}&\geq& (1+\delta)^{(t-1)/\delta}\left[F(\mathbf{x}_t\vee\mathbf{1}_{OPT})+F(\mathbf{x}_{t+\delta})-F(\mathbf{x}_t)\right]\\
	&&+~ \sum_{o\in OPT}\ell(o)(1-\mathbf{x}_t(o)) -5\epsilon\tau.
\end{eqnarray*}
An observation is that this derivative is determined by $F(\mathbf{x}_t\vee\mathbf{1}_{OPT})$, $F(\mathbf{x}_{t+\delta})$ and $F(\mathbf{x}_t)$. 
However, these terms are not easily eliminated in the process of telescoping sum due to the adaptive weights as a function of the timestamp $t$.
The next lemma improves this dilemma and gives a more straightforward lower bound comparing with Lemma \ref{update_profit} on the increase of each update by releasing some of the redundant terms at a small cost.

\begin{lemma}
	\label{update_profit_final}
	With probability $1-\mathcal{O}(\frac{1}{n})$, the following holds
	\begin{eqnarray*}
		\dfrac{\Gamma(t+\delta)-\Gamma(t)}{\delta}\geq(1+\delta)^{(t-1)/\delta}\cdot F(\mathbf{x}_t\vee\mathbf{1}_{OPT})+ \sum_{o\in OPT}\ell(o)(1-\mathbf{x}_t(o)) -8\epsilon\tau,
	\end{eqnarray*}
	where $\tau=\max_{s\in\mathcal{S}}f(s)$.
\end{lemma}
\begin{proof}
	We condition on the event $\Xi$ for the rest of the proof. Due to Lemma \ref{update_profit} and the local nearly linearity of multilinear extension (stated as Lemma \ref{multilinear_locallinear}), we have
	\begin{eqnarray*}
		\dfrac{\Gamma(t+\delta)-\Gamma(t)}{\delta}
		&\geq& (1+\delta)^{(t-1)/\delta}[F(\mathbf{x}_{t+\delta}) - F(\mathbf{x}_t)] - 5\epsilon \tau \\
		&&+~ (1+\delta)^{(t-1)/\delta}F(\mathbf{x}_t\vee\mathbf{1}_{OPT}) + \sum_{o\in OPT}\ell(o)(1-\mathbf{x}_t(o))  \\
		&\geq& (1+\delta)^{(t-1)/\delta}\left[\langle \delta \mathbf{z}_t\odot(1-\mathbf{x}_t), w_{\mathbf{x}_t}\rangle - 3\epsilon\delta \tau\right] - 5\epsilon \tau \\
		&&+~ (1+\delta)^{(t-1)/\delta}F(\mathbf{x}_t\vee\mathbf{1}_{OPT}) + \sum_{o\in OPT}\ell(o)(1-\mathbf{x}_t(o)) \\
		&\geq& (1+\delta)^{(t-1)/\delta}F(\mathbf{x}_t\vee\mathbf{1}_{OPT}) + \sum_{o\in OPT}\ell(o)(1-\mathbf{x}_t(o)) \\
		&& -~ (5+3(1+\delta)^{(t-1)/\delta}\delta)\epsilon \tau \\
		&\geq&(1+\delta)^{(t-1)/\delta}F(\mathbf{x}_t\vee\mathbf{1}_{OPT}) + \sum_{o\in OPT}\ell(o)(1-\mathbf{x}_t(o)) - 8\epsilon \tau,
	\end{eqnarray*}
	where the last inequality follows from $(1+\delta)^{(t-1)/\delta}\leq1$, since $(t-1)/\delta\leq0$, and the third inequality is given by the truth that the term $\langle \mathbf{z}_t\odot(1-\mathbf{x}_t), (1+\delta)^{(t-1)/\delta}\cdot w_{\mathbf{x}_t}+ \ell \rangle$ is non-negative, i.e.,
	$$\langle \mathbf{z}_t\odot(1-\mathbf{x}_t), (1+\delta)^{(t-1)/\delta}\cdot w_{\mathbf{x}_t}\rangle \geq\langle \mathbf{1}_{\emptyset}\odot(1-\mathbf{x}_t), (1+\delta)^{(t-1)/\delta}\cdot w_{\mathbf{x}_t}\rangle =0,$$
	where the inequality is guaranteed since $f$ is non-negative, $\mathcal{P}$ is down-monotone and $\mathbf{1}_{\emptyset}$ is one possible option for the choice of $\mathbf{z}_t$.
	\qed
\end{proof}

According to the analysis routine of the greedy-like methods in submodular maximization problem, we should lower bound the guarantee of sum $F(\mathbf{x}_1)+L(\mathbf{x}_1)$ of the output $\mathbf{x}_1$ by a final telescoping sum with respect to the time interval, after obtaining the improved derivative.

\begin{theorem}
	\label{thm}
	The algorithm \textsc{Measured Continuous Greedy with Adaptive Weights} outputs $\mathbf{x}_1\in\mathcal{P}$ and with probability $1-\mathcal{O}(\frac{1}{n})$, it satisfies 
	$$F(\mathbf{x}_1)+L(\mathbf{x}_1)\geq (\frac{1}{e}-\epsilon)\cdot f(OPT)+\left(\frac{\beta-e}{e(\beta-1)}-\mathcal{O}(\epsilon)\right)\cdot\ell(OPT)-8\epsilon\tau,$$
	where $\beta\in[0,\infty]$, $\tau=\max_{s\in\mathcal{S}}f(s).$
\end{theorem}
\begin{proof}
	First, $\mathbf{x}_1\in\mathcal{P}$ since the polytope is down-monotone and $\mathbf{x}_1$ is a convex combination of $\delta^{-1}$ vectors $\mathbf{z}_t\odot(1-\mathbf{x}_t)$ in $\mathcal{P}$.
	Next, from Lemma \ref{func_esti}, we know the event $\Xi$ happens with probability $1-\mathcal{O}(\frac{1}{n})$. Consider the output $\mathbf{x}_1$ of the Algorithm \ref{MCGAW}, Lemma \ref{update_profit_final} guarantees the inequality below
	\begin{eqnarray*}
		&&F(\mathbf{x}_1)+L(\mathbf{x}_t) \\
		&=&\Gamma(1)=\Gamma(0)+\sum_{t\in \mathcal{G}}[\Gamma(t+\delta)-\Gamma(t)] \\
		&\geq& \Gamma(0)+\delta\sum_{t\in\mathcal{G}}\left[ (1+\delta)^{(t-1)/\delta}\cdot F(\mathbf{x}_t\vee\mathbf{1}_{OPT})+\sum_{o\in OPT}\ell(o)(1-\mathbf{x}_t(o))-8\epsilon\tau \right].
	\end{eqnarray*}
	
	Notice that if we want to build a connection between the term $F(\mathbf{x}_t\vee\mathbf{1}_{OPT})$ and the optimal value $f(OPT)$, we have to get rid of the current continuous solution $\mathbf{x}_t$. Fortunately, we could replace it with the discrete value $f(OPT)$ by the property of multilinear extension (stated as Lemma \ref{multilinear_lowerbound}). Specifically, we get
	$$F(\mathbf{x}_t\vee\mathbf{1}_{OPT})\geq(1-\theta)\cdot f(OPT),$$
	where $\theta$ is an upper bound with $\mathbf{x}_s\leq\theta$ for every $s\in\mathcal{S}$. 
	Besides, we also need the following claim to find a union bound for $\theta$. We omit the proof here and it can be found in \cite{FNS11}.
	
	\begin{claim}
		\label{tech_claim}
		For every $t\in\mathcal{G}$ and $\delta$ in the algorithm \textsc{Measured Continuous Greedy with Adaptive Weights}, the inequality below is true for any $s\in\mathcal{S}$, $$\mathbf{x}_t(s)\leq 1-(1-\delta)^{t/\delta} \leq 1-e^{-t}+\mathcal{O}(\delta)=1-e^{-t}+\epsilon.$$
	\end{claim}
	
	Therefore, for the term $F(\mathbf{x}_t\vee\mathbf{1}_{OPT})$, we obtain 
	$$F(\mathbf{x}_t\vee\mathbf{1}_{OPT})\geq(e^{-t}-\epsilon)\cdot f(OPT).$$
	
	We apply a similar trick to deal with the term $\sum_{o\in OPT}\ell(o)(1-\mathbf{x}_t(o))$. Firstly, We denote sets $OPT^+=\{o|\ell(o)\geq 0, o\in OPT\}$ and $OPT^-=\{o|\ell(o)< 0, o\in OPT\}$, respectively. Then, 	
	\begin{eqnarray*}
		\sum_{o\in OPT}\ell(o)(1-\mathbf{x}_t(o)) 
		&=& \sum_{o\in OPT^+}\ell(o)(1-\mathbf{x}_t(o)) + \sum_{o\in OPT^-}\ell(o)(1-\mathbf{x}_t(o)) \\
		&\geq& (e^{-t}-\epsilon) \sum_{o\in OPT^+}\ell(o) + \sum_{o\in OPT^-}\ell(o) \\
		&\geq& (\frac{1}{e}-\epsilon) \sum_{o\in OPT^+}\ell(o) + \sum_{o\in OPT^-}\ell(o),
	\end{eqnarray*}	
	where the first inequality holds due to the claim above and the definition of $OPT^-$ and the second one is true since $t\leq 1$. 
	
	A natural thought about the optimal solution $OPT$ is that its positive and negative parts, i.e., $OPT^+$ and $OPT^-$, must satisfy a certain ratio and we assume that it is $\beta=\sum_{o\in OPT^+}\ell(o)/|\sum_{o\in OPT^-}\ell(o)|\in[0,\infty]$. Now we can easily have
	$$(\frac{1}{e}-\epsilon) \sum_{o\in OPT^+}\ell(o) + \sum_{o\in OPT^-}\ell(o) = \left(\frac{\beta-e}{e(\beta-1)}-\mathcal{O}(\epsilon)\right)\cdot\ell(OPT).$$
	
	Combining with the above result we can get
	\begin{eqnarray*}
		&& \Gamma(0)+\delta\sum_{t\in\mathcal{G}}\left[ (1+\delta)^{(t-1)/\delta}\cdot F(\mathbf{x}_t\vee\mathbf{1}_{OPT})+\sum_{o\in OPT}\ell(o)(1-\mathbf{x}_t(o))-8\epsilon\tau \right] \\
		&\geq& \frac{1}{e} f(\emptyset) + f(OPT)\cdot\sum_{t\in\mathcal{G}} \delta(1+\delta)^{(t-1)/\delta}\left(e^{-t}-\epsilon\right) - 8\epsilon\tau \\
		&&+~   \left(\frac{\beta-e}{e(\beta-1)}-\mathcal{O}(\epsilon)\right)\cdot\ell(OPT),
	\end{eqnarray*}
	where the inequality follows from $|\mathcal{G}|=\delta^{-1}$ and $\Gamma(0)=(1+\delta)^{-1/\delta}\cdot f(\emptyset)+\ell(\emptyset)=(1+\delta)^{-1/\delta}\cdot f(\emptyset)\geq 1/e\cdot f(\emptyset)$, since $f(S)\geq 0$ for each $S\subseteq\mathcal{S}$ and $\ell$ is additive.

	Therefore, the inequality of $F(\mathbf{x}_1)+\ell(\mathbf{x}_1)$ can be further derived to the following result
	\begin{eqnarray*}
		F(\mathbf{x}_1)+\ell(\mathbf{x}_1)
		&\geq& \frac{1}{e} f(\emptyset) + f(OPT)\cdot\sum_{t\in\mathcal{G}} \delta(1+\delta)^{(t-1)/\delta}\left(e^{-t}-\epsilon\right) \\
		&& +~ \left(\frac{\beta-e}{e(\beta-1)}-\mathcal{O}(\epsilon)\right)\cdot\ell(OPT) - 8\epsilon\tau.
	\end{eqnarray*}
	
	Now the only task of the proof we have left is to give a lower bound to the telescoping sum. Since the timestamp $t$ moves forward with a constant interval $\delta$, we get
	\begin{eqnarray*}
		\sum_{t\in\mathcal{G}}\delta(1+\delta)^{(t-1)/\delta}&=&\sum_{i=0}^{\delta^{-1}-1}\delta(1+\delta)^{(i\delta-1)/\delta}=\delta(1+\delta)^{-\delta^{-1}}\cdot\sum_{i=0}^{\delta^{-1}-1}(1+\delta)^i \\
		&=&\delta(1+\delta)^{-\delta^{-1}}\cdot\dfrac{1-(1+\delta)^{\delta^{-1}}}{1-(1+\delta)}=1-(1+\delta)^{-\delta^{-1}} \\
		&\geq& 1-\frac{1}{e(1-\delta)}\geq 1-\frac{1+2\delta}{e}\geq 1-1/e-\frac{\epsilon}{en},
	\end{eqnarray*}
	where the three inequalities are guaranteed since $(1+\frac{1}{\gamma})^{\gamma}\geq e(1-\frac{1}{\gamma})$ for every $\gamma\geq 1$, $\frac{1}{1-\gamma}\leq 1+2\gamma$ for every $\gamma\leq\frac{1}{2}$ and $\delta\leq\frac{\epsilon}{2n^2}$, respectively.
	
	Moreover, for the term $\sum_{t\in\mathcal{G}}\left(e^{-t}-\epsilon\right)$, it satisfies 
	$$\sum_{t\in\mathcal{G}}\left(e^{-t}-\epsilon\right)\geq \frac{1}{e-1}-\epsilon.$$
	
	Then, we get the following targeted constant lower bound for the telescoping sum
	$$\sum_{t\in\mathcal{G}}\delta(1+\delta)^{(t-1)/\delta}\left(e^{-t}-\epsilon\right) \geq (1-\frac{1}{e}-\frac{\epsilon}{en})\cdot\left(\frac{1}{e-1}-\epsilon\right) = \frac{1}{e}-\epsilon.$$
	
	Combining all the results above, we finally obtain the conclusion
	$$F(\mathbf{x}_1)+\ell(\mathbf{x}_1)\geq (\frac{1}{e}-\epsilon)\cdot f(OPT)+\left(\frac{\beta-e}{e(\beta-1)}-\mathcal{O}(\epsilon)\right)\cdot\ell(OPT)-8\epsilon\tau.$$
	\qed
\end{proof}

\section{Additional Results}
\label{sec_addtionalresults}
Some additional results are presented in this section. For the completeness of this work, we first submit the theoretical version of Algorithm \ref{MCGAW} in continuous time setting. Then, we propose the result of applying our algorithm in the case of monotone submodular function.

\subsection{Theoretical Algorithm}
The continuous version of our algorithm is given below. It shows the running process from a continuous time perspective.

\begin{algorithm}[htbp]
	\caption{\textsc{Measured Continuous Greedy with Adaptive Weights (continuous)}}
	\label{MCGAW_continuous}
	\begin{algorithmic}[1]
		\REQUIRE submodular function $f$, modular vector $\vec{\ell}$, polytope $\mathcal{P}$, parameter $\epsilon$.
		\ENSURE $\mathbf{x}_1$
		\STATE Initialize: $\mathbf{x}_0\leftarrow\mathbf{1}_{\emptyset}$.
		\WHILE {$t \in [0,1)$}
		\FOR {$s\in\mathcal{S}$}
		\STATE $w_{\mathbf{x}_t}(s)\leftarrow\mathbb{E}[f_{R_{\mathbf{x}_t}-s}(s)]$.
		\ENDFOR
		\STATE $\mathbf{z}_{t}\leftarrow\arg\max_{\mathbf{z}_{t}\in \mathcal{P}}\left\langle \mathbf{z}_{t}\odot(1-\mathbf{x}_{t}), e^{t-1}\cdot w_{\mathbf{x}_t}+ \vec{\ell} \right\rangle$.
		\STATE $\frac{d\mathbf{x}_t}{dt}\leftarrow \mathbf{z}_{t}\odot(1-\mathbf{x}_{t})$.
		\ENDWHILE
	\end{algorithmic}
\end{algorithm}

The main difference between the discrete algorithm (Algorithm \ref{MCGAW}) and the ideal algorithm (Algorithm \ref{MCGAW_continuous}) is that we do not have to do sampling to estimate the marginal contribution in each dimension of the vector $\mathbf{x}_t$. Therefore, we can precisely use the gradient information to guide the update direction. After a similar derivation process in Section \ref{sec_analysis}, it is naturally to get the results below.

\begin{theorem}
	\label{thm_continuous}
	For the problem of maximizing $F(\mathbf{x})+\ell(\mathbf{x})$ over a down-closed polytope $\mathcal{P}$, the \textsc{Measured Continuous Greedy with Adaptive Weights (continuous)} outputs $\mathbf{x}_1\in\mathcal{P}$ and it satisfies 
	$$F(\mathbf{x}_1)+L(\mathbf{x}_1)\geq (\frac{1}{e}-\epsilon)\cdot f(OPT)+ \left(\frac{\beta-e}{e(\beta-1)}-\mathcal{O}(\epsilon)\right)\cdot \ell(OPT).$$
\end{theorem}

Moreover, we could also give a more general result inspired by the idea in \cite{Feldman21}.

\begin{theorem}
	\label{thm_continuous_general}
	For maximizing $F(\mathbf{x})+\ell(\mathbf{x})$ over a down-monotone polytope $\mathcal{P}$, the \textsc{Measured Continuous Greedy with Adaptive Weights (continuous)} outputs $\mathbf{x}_1\in\mathcal{P}$ and it satisfies 
	$$F(\mathbf{x}_1)+L(\mathbf{x}_1)\geq \max_{\lambda\in[0,1]}(\frac{\lambda}{e}-\epsilon)\cdot f(OPT)+\max_{\lambda\in[0,1]}\left(\frac{\lambda(\beta-e)}{e(\beta-1)}-\mathcal{O}(\epsilon)\right)\cdot\ell(OPT).$$
\end{theorem}

\subsection{Monotone Case}
When the submodular function in the sum is monotone, our algorithms (Algorithm \ref{MCGAW} and Algorithm \ref{MCGAW_continuous}) still work. Despite the algorithms take more steps and update slowly within a factor $(1-\mathbf{x}_t)$, we can achieve the same ratio as Feldman did in \cite{Feldman21} as long as we do a similar analysis process like the last section. For this reason, we omit the proofs and just provide the results here.

\begin{theorem}
	\label{thm_monotone}
	When $f$ is monotone, both Algorithm \ref{MCGAW} and Algorithm \ref{MCGAW_continuous} can produce a vector $\mathbf{x}_1\in\mathcal{P}$ and with probability $1-\mathcal{O}(\frac{1}{n})$ it satisfies 
	$$F(\mathbf{x}_1)+L(\mathbf{x}_1)\geq \max_{\lambda\in[0,1]}(1- e^{-\lambda}-\epsilon)\cdot f(OPT)+\max_{\lambda\in[0,1]}\left(\frac{\lambda(\beta-e)}{e(\beta-1)}-\mathcal{O}(\epsilon)\right)\cdot\ell(OPT).$$
\end{theorem}

\section{Inapproximability Results}
\label{sec_hardness}
\begin{algorithm}[htbp]
	\caption{Reduction \cite{Feldman21}}
	\label{Reduction}
	\begin{algorithmic}[1]
		\REQUIRE submodular function $f$, an instance $\mathcal{I}$ in the down-closed family of sets with rank $k$, a polynomial algorithm $M$.
		\ENSURE set $S$
		\STATE Denote function $\ell$ by $\ell(S)=\frac{m}{k}\cdot|S|$ for every set $S\in\mathcal{I}$.
		\STATE Run $M$ with input $f$ and $\ell$.
	\end{algorithmic}
\end{algorithm}

About the inapproximability results for submodular maximization, Gharan and Vondr\'ak \cite{GV11} presented a result by simulated annealing. They showed that we can not get a better solution with $0.478$-approximation guarantee for maximizing a submodular set function over a matroid independence constraint. Formally, their conclusion can be stated as the following proposition.

\begin{proposition}[\cite{GV11}]
	\label{hardness_matroid}
	Given the value oracle of the submodular set function $f:2^{\mathcal{S}}\rightarrow\mathbb{R}_{\geq 0}$, we can only find a feasible solution for maximizing $f$ over a matroid constraint with $0.478$-approximation in at least exponentially many value queries.
\end{proposition}

Since the matroid constraint is included in the down-closed family of sets, we can also give a result from hardness side for maximizing the sums of submodular and modular with the help of Proposition \ref{hardness_matroid}. The high level intuition is to show a violation of Proposition \ref{hardness_matroid}, which means that for a given matroid $\mathcal{M}$, we could obtain an instance of $\max\{f(S):S\in\mathcal{M}\}$ in polynomial time. To do so, we assume that there is a polynomial-time algorithm $M$, whose output $S\in\mathcal{I}$ satisfies the target guarantee $f(S)+\ell(S)\geq 0.478\cdot f(OPT)+\lambda\ell(OPT)$, where $\lambda\in[0,1]$. Then, we directly inherit the simple \textsc{Reduction} algorithm (stated as Algorithm \ref{Reduction}) in \cite{Feldman21} with a slightly change in the given constraint and construct the contradiction through it. Now we state the hardness result formally.

\begin{theorem}
	\label{hardness_SumSubMod}
	For the problem $\max_{S\in\mathcal{I}}\{f(S)+\ell(S)\}$, where $f: 2^{\mathcal{S}}\rightarrow\mathbb{R}_{\geq 0}$ is submodular, $\ell: 2^{\mathcal{S}}\rightarrow\mathbb{R}$ is modular and $\mathcal{I}$ is the given down-closed family of sets, there exist no polynomial time algorithm whose output $S$ satisfies $f(S)+\ell(S)\geq\max_{\lambda\in[0,1],T\subseteq\mathcal{I},|T|\leq k}\{0.478\cdot f(T)+\lambda\ell(T)\}$, where $k$ denote the size of the maximal feasible set in $\mathcal{I}$.
\end{theorem}
\begin{proof}
	To construct a contradiction we consider a special case that $f$ and $\ell$ satisfy $f(S)/\ell(S)=m^{-1}$ for every $S\in\mathcal{I}$.
	For ease of explanation, we assume that the value of $OPT$ to the submodular function is $1$. Hence, a obvious observation is $\ell(OPT)=m$ since $f(OPT)=1$ and $m\cdot f(OPT)=\ell(OPT)$.
	
	Therefore, by the definition of $\ell$ in Algorithm \ref{Reduction} the output set $S$ satisfies
	$$f(S)=f(S)+\ell(S)-\frac{m}{k}\cdot|S|.$$
	
	Then, by our assumption about the polynomial time algorithm $M$, we have
	$$f(S)+\ell(S)\geq0.478\cdot f(OPT)+\lambda\ell(OPT),$$
	where $\lambda\in[0,1]$.
	
	Combining the two inequalities above, we can obtain
	\begin{eqnarray*}
		f(S)
		&\geq&0.478\cdot f(OPT)+\lambda\ell(OPT)-\frac{m}{k}\cdot|S| \\
		&=&0.478\cdot f(OPT)+\lambda m-\frac{m}{k}\cdot|S| \\
		&=&0.478\cdot f(OPT)+m(\lambda-\frac{|S|}{k}).
	\end{eqnarray*}

	By plugging $\lambda=\frac{|S|}{k}\in[0,1]$, we finally get $f(S)\geq0.478\cdot f(OPT)$, which violates Proposition \ref{hardness_matroid}. \qed
\end{proof}

\section{Conclusion}
\label{sec_conclusion}
In the paper, we give an approximation algorithm for maximizing non-monotone multilinear extension plus modular functions over a down-monotone polytope, which is a general form containing several common constraints such as matroid and knapsack. The solution given by the algorithm with high probability satisfies the guarantee $F(\mathbf{x})+L(\mathbf{x})\geq (1/e-\epsilon)\cdot f(OPT)+\left(\frac{\beta-e}{e(\beta-1)}-\mathcal{O}(\epsilon)\right)\cdot\ell(OPT)$ with $\mathcal{O}(\epsilon)$ additive error. The result is obtained by a trivial intuition that we can use a parameter $\beta$ to portray the relative relationship between the non-negative part w.r.t. the optimal modular value and its negative part. Since we do not make any assumption to the value range of the modular function, our result is quiet general so that it can reduce to many existed results. For example, the above guarantee reduces to Lu et al.'s result \cite{LYG21}, i.e., the guarantee of the modular term is $1$, when the negative part is completely dominant ($\beta=0$). Moreover, the algorithm also works for the case that the submodular function in the sums is monotone. Besides, an inapproximability for the original problem we consider is provided. It tells that we cannot find a feasible solution $S$ satisfying $f(S)+\ell(S)\geq\max_{\lambda\in[0,1],T\subseteq\mathcal{I},|T|\leq k}\{0.478\cdot f(T)+\lambda\ell(T)\}$ in polynomial time, where the constraint $\mathcal{I}$ is an instance of the down-closed family of sets and $k$ denotes the size of maximal feasible sets in $\mathcal{I}$.

\section{Acknowledgments}
The authors sincerely thank Dr. M. Feldman for conversations that helped this work.

\end{document}